# MCPA: Multi-scale Cross Perceptron Attention Network for 2D Medical Image Segmentation

Liang Xu[1], *Mingxiao Chen[1], Yi Cheng*, Pengfei Shao, Shuwei Shen[*], Peng Yao[*], and Ronald X.Xu[*]

**Abstract**—The UNet architecture, based on Convolutional Neural Networks (CNN), has demonstrated its remarkable performance in medical image analysis. However, it faces challenges in capturing long-range dependencies due to the limited receptive fields and inherent bias of convolutional operations. Recently, numerous transformer-based techniques have been incorporated into the UNet architecture to overcome this limitation by effectively capturing global feature correlations. However, the integration of the Transformer modules may result in the loss of local contextual information during the global feature fusion process. To overcome these challenges, we propose a 2D medical image segmentation model called Multi-scale Cross Perceptron Attention Network (MCPA). The MCPA consists of three main components: an encoder, a decoder, and a Cross Perceptron. The Cross Perceptron first captures the local correlations using multiple Multi-scale Cross Perceptron modules, facilitating the fusion of features across scales. The resulting multi-scale feature vectors are then spatially unfolded, concatenated, and fed through a Global Perceptron module to model global dependencies. Furthermore, we introduce a Progressive Dual-branch Structure to address the semantic segmentation of the image involving finer tissue structures. This structure gradually shifts the segmentation focus of MCPA network training from large-scale structural features to more sophisticated pixel-level features. We evaluate our proposed MCPA model on several publicly available medical image datasets from different tasks and devices, including the open large-scale dataset of CT (Synapse), MRI (ACDC), fundus camera (DRIVE, CHASE_DB1, HRF), and OCTA (ROSE). The experimental results show that our MCPA model achieves state-of-the-art performance. The code is available at https://github.com/simonustc/MCPA-for-2D-Medical-Image-Segmentation.

## I. INTRODUCTION

IN recent years, the rapidly evolving medical imaging technologies have played a critical role in the diagnosis and treatment of disease. Modalities such as computed tomography (CT), magnetic resonance imaging (MRI), fundus camera, and optical coherence tomography angiography (OCTA) have become primary clinical examinations in hospitals [1]. These imaging modalities provide valuable structural and functional information about organs and surrounding tissues [2]-[6]. They also help to distinguish between different organs and tissues, which is particularly important for accurate functional diagnosis. However, even experienced physicians require significant manual effort for image segmentation, which can be labor-intensive and error-prone. Accurate and efficient segmentation of medical images has therefore become an essential step in clinical applications.

Among the existing medical image segmentation techniques, the typical U-shaped network(U-Net) has achieved remarkable success [7]. By incorporating skip connections between the encoder and decoder layers, U-Net effectively addresses the spatial information loss caused by downsampling operations. Recent advances in this architecture have introduced variants with more efficient CNN structures and denser nested skip connections, such as Res-UNet [8], Eff-UNet [9], Unet++ [10], nnU-net [11], and others. Although CNN-based UNet models have shown promising results, it is important to recognize their limitations in capturing global context and long-range spatial dependencies, as they can only extract local features using fixed-size convolution kernels [12]-[16].

Recently, transformers that efficiently model global dependencies via self-attention mechanisms have been integrated into the UNet architecture and have shown impressive performance [17]-[21]. TransUNet [15] was a pioneering work that introduced a complete Vision Transformer (ViT) structure after the last layer of the UNet encoder, and explored the potential of Transformers in medical image segmentation. Subsequent Transformer-based UNets such as CoTr [13], SpecTr [22], SwinUnet [14], MISSFormer [16], nnFormer [23], C2FTrans [24], and others, have also demonstrated excellent performance. MISSFormer [16] attempts to capture local and global correlations of some multi-scale features through an enhanced transformer context bridge based on enhanced transformer blocks. C2FTrans [24] uses a cross-scale global transformer (CGT) to extract global scale dependencies with low computational complexity. However, despite the different strategies employed by these methods to address the shortcomings of Transformers in handling local details, they still fall short of complete success, leaving room for further improvement.

1 indicates co-first authorship. Asterisk indicates corresponding author.

L. Xu, S. Shen, and R. Xu are with the Suzhou Institute for Advanced Research, University of Science and Technology of China (USTC), Suzhou 215123, China. E-mail: xul666@mail.ustc.edu.cn, {swshen, xux}@ustc.edu.cn.

Mingxiao Chen, Yi. Cheng, Pengfei Shao, are with the Department of Precision Machinery and Precision Instrument, USTC, Hefei 230026, China. E-mail: {cmx173, mchengyi}@mail.ustc.edu.cn, spf@ustc.edu.cn.

Peng. Yao are with the School of Microelectronics, USTC, Hefei 230026, China. E-mail: yaopeng@ustc.edu.cn.

Furthermore, for tasks involving the segmentation of small tissue structures, such as retinal segmentation with intricate fine blood vessels, it is essential to improve the network's focus on accurately delineating these fine structures [25]-[29]. Specifically, Li *et al.* [30] proposed a Response Cue Erasing (RCE) module that erases positions in the input image corresponding to confident pixels in the output of the main branch, thereby further improving segmentation accuracy. In addition, Ma *et al.* [29] introduced OCTA-Net, a novel split-based coarse-to-fine vessel segmentation network designed for OCTA images. OCTA-Net is capable of separately detecting thick and thin vessels. It is noteworthy that existing works dealing with the segmentation of such complex structures predominantly rely on CNN architectures. This preference may be due to the fact that transformers have received comparatively less attention in this field due to their perceived limitations in effectively handling local details.

In this paper, we propose a 2D medical image segmentation network called the Multi-scale Cross Perceptron Attention Net (MCPA). MCPA consists of an encoder, a decoder, and a Cross Perceptron. The Cross Perceptron establishes local correlations between features at different scales and then models global dependencies in the spatial dimension, thereby achieving the fusion of features at different scales. To address the challenge of segmenting fine tissue structures, we introduce a Progressive Dual-branch Structure (PDBS). This structure progressively shifts the focus of the network from the segmentation results of the whole image to the finer pixel-level features. Our main contributions to this work are summarized below:

- We propose a novel MCPA network that incorporates a Cross Perceptron to achieve local correlation and global dependence, effectively fusing features of different scales.
- We introduce a Progressive Dual-branch Structure (PDBS), which progressively shifts the training focus of the network to finer pixel-level features while incorporating the cross perceptron, resulting in improved segmentation accuracy.
- We evaluate the performance of MCPA on various datasets from different medical imaging modalities, including CT (Synapse), MRI (ACDC), retinal fundus camera (DRIVE, CHASE_DB1, HRF), and OCTA (ROSE). The results show that MCPA outperforms existing 2D medical image segmentation models based on CNN and Transformer architectures.

## II. METHODS

For the medical image segmentation task, we propose a novel network called Multi-scale Cross Perceptron Attention Network (MCPA). As shown in Fig. 1, the network consists of an encoder, a decoder, and a Cross Perceptron (as shown in Fig. 2). The encoder is responsible for extracting rich semantic features from the input image, while the decoder generates the segmentation mask. Unlike the original UNet we propose the Cross Perceptron to replace the independent skip connections between corresponding layers in the encoder and decoder. This design aims to capture the long-range dependencies between different scales of features, allowing the network to effectively integrate information from multiple scales.

Furthermore, to address the challenging task of semantic segmentation in medical images that containing fine tissue structures, such as retinal vessel segmentation in fundus images, we propose a Progressive Dual-branch Structure (PDBS). The presented innovative architecture, as shown in Fig. 3, consists of two parallel branches based on the MCPA. The Main branch aims to achieve coarse-grained overall segmentation, while the Fine branch focuses on fine-grained detailed segmentation. To establish communication between the two branches, we introduce the RCE module [30]. During training, a Progressive Regularization Loss (PR Loss, $L_P$) is proposed to gradually shift the training emphasis from the Main branch to the Fine branch, effectively improving the segmentation performance in the presence of intricate details.

### A. Multi-scale Cross Perceptron Attention Net

#### 1) Encoder and Decoder

To ensure a concise overall architecture, we adopt the Shunted Self-Attention (SSA) network [30] as the backbone in both the encoder and decoder of our proposed network. The traditional transformer calculates Q, K, and V through the self-attention mechanism, but the three scales are the same [17]. While SSA provides K and V of different scales through Multi-Scale Token Aggregation (MTA), and also adaptively merge tokens on large objects and retain tokens on small objects. Both the encoder and the decoder consist of four Stages, each of which consists of a Patch Embedding or a Patch Expand operation, along with several SSA Blocks (in Stage 4 of the decoder, the SSA Block is replaced by Linear Projection).

The input image $x \in \mathbb{R}^{H \times W \times C}$ is first downsampled and reshaped into a sequence of flattened 2D patches $x_P$ using the Patch Embedding operation. In Stage 1, the Patch Embedding operation uses three convolutional layers with kernel sizes of 7, 3, 2, and strides of 2, 1, 2, respectively. This results in a 4-fold reduction in spatial dimensions compared to the input image. In Stages 2-4, only one convolution layer with a kernel size of 3 is used, and feature maps are downsampled 2-fold in each Stage.

Similarly, in the decoder from Stage 1 to 4, the Patch Expand operation in the deconvolution layers gradually restores the resolution of the feature map to its original dimensions. However, there is an additional step involved. In Decoder Stages 1-3, the outputs of Patch Expand in each Stage are concatenated with features generated by the Cross Perceptron module. After concatenation, the resulting feature map is passed through a Linear layer to halve the number of channels before being fed into the SSA Block.

As shown in Fig. 1(b), the SSA Block consists of SSA Attention and SSA Feed Forward Network (SSA FFN). The SSA Attention uses Multi-Scale Token Aggregation (MTA) to aggregate information from different scales. The resulting tensors are then concatenated and used to compute the attention scores, as shown in (1):

$$
\begin{aligned}
Q_i &= XW_i^Q, \\
K_i, V_i &= MTA(X, r_i)W_i^K, MTA(X, r_i)W_i^V, \\
V_i &= V_i + DW(V_i; \theta_1),
\end{aligned}
\tag{1}
$$

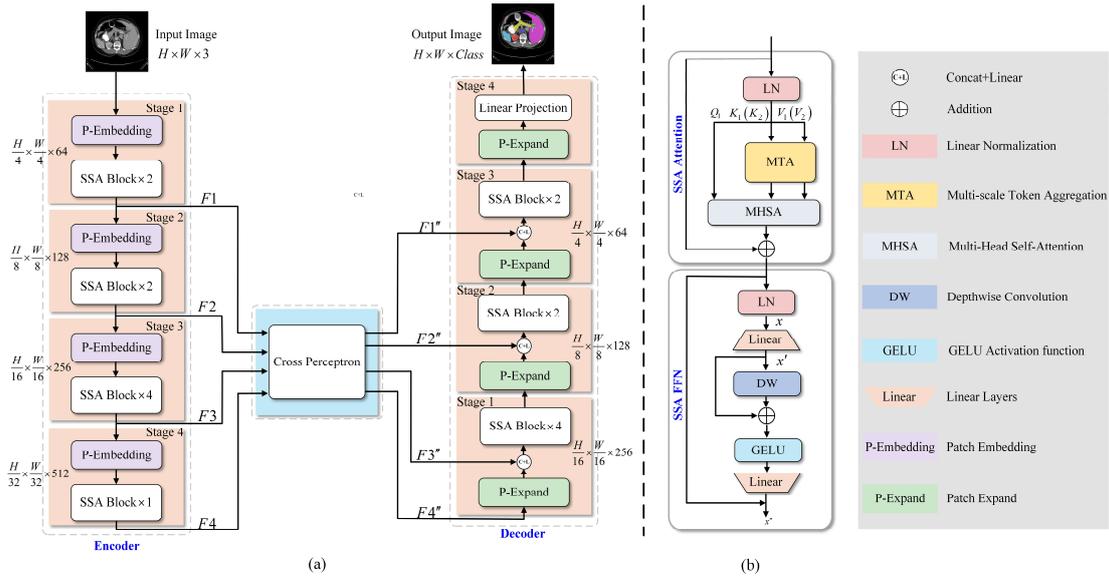

Fig 1. (a) Overall architecture of the proposed Multi-scale Cross Perceptron Attention Net (MCPA). (b) Details of the SSA Block.

where $MTA(; r_i)$ is the Multi-scale Token Aggregation layer in the i-th head with the downsampling rate of $r_i$ by using a convolutional layer. $DW()$ is a Depthwise Convolution layer with parameters $\theta_1$. As $r_i$ increases, more tokens in $K$ and $V$ are merged, resulting in shorter lengths for $K$ and $V$, preserving the ability to capture larger objects.

Then, the SSA Attention output is used as the input of SSA FFN, as shown in (2):

$$x' = Linear(x);$$
$$x'' = Linear(\sigma(x' + DW(x'; \theta_2))), \quad (2)$$

where $Linear()$ is a Linear layer, and $\sigma()$ is the GELU activation function. For a detailed description of SSA, please refer to [21].

Finally, we adopt the Cross Entropy Loss and Dice Loss [31] as the segmentation loss, which quantify the dissimilarity between the ground truth labels and the model predictions. The combined loss function is defined as follows:

$$L_{total} = \lambda L_{CE} + (1-\lambda)L_{Dice}, \quad (3)$$

where $\lambda$ is a weight hyperparameter and fixed to 0.4.

### 2) Cross Perceptron

To effectively integrate multi-scale information, we propose the Cross Perceptron for feature fusion across different scales. As shown in Fig. 2(a), we use four Cross Perceptron modules, namely Perceptron 1 to Perceptron 4, to capture local correlations between features at different scales. This allows the extracted features to contain richer semantic information as well as more local detail information. To facilitate global dependency modeling, following the approach in [16], we use a Global Perceptron (G-Perceptron) module to unfold and concatenate the multi-scale feature vectors across the spatial dimensions.

Fig. 2(b) shows the architecture of the Cross Perceptron Attention (CPA) module, which serves as the core module of Perceptron 1 to Perceptron 3. The CPA module accepts two inputs, one is from the corresponding Stage output of the encoder, and the other is sourced from either a lower Stage or a Perceptron. Both inputs are individually passed through a Linear Layer, resulting in $Q$, $K$, and $V$ projections. These projections are then fed into a Multi-Head Cross-Attention (MHCA), which calculates the cross-attention values between $Q$, $K$, and $V$ from different sources. The calculation formula remains consistent with the classical Self-Attention formula :

$$Attention(Q, K, V) = softmax(\frac{QK^T}{\sqrt{d_k}})V, \quad (4)$$

where $d_k$ is the dimension of the attention heads, $K^T$ is the transpose of $K$.

Note that, the inputs in this study to the CPA module are derived from encoders at different Stages or other Multi-scale Cross Perceptron modules, resulting in different sequence lengths. Consequently, it is necessary to adjust the sequence length of the output features $Q$, $K$, and $V$ so that the cross-attention values can be computed (as described in (4)). To strike a balance between computational complexity and accuracy, we set specific sequence lengths for $Q$, $K$, and $V$ within each CPA module (Perceptron 1 to Perceptron 3), as shown in Table I. Note that the lengths of $Q$ remain consistent across CPAs within the same Multi-scale Cross Perceptron module, while the lengths of $K$ and $V$ can be either the same (e.g., Perceptron 1) or different (e.g., Perceptron 2, Perceptron 3).

Within each Multi-scale Cross Perceptron module, the output of multiple CPAs are concatenated and then subsequently passed through a Linear Layer to reduce their dimensionality. This dimensionality reduction aligns the output with the original output signal dimensionality of the corresponding stage in the encoder. The resulting output is the further processed by the FFN module, as shown in Fig. 2(d). Mathematically, the entire FFN operation can be expressed as follows:

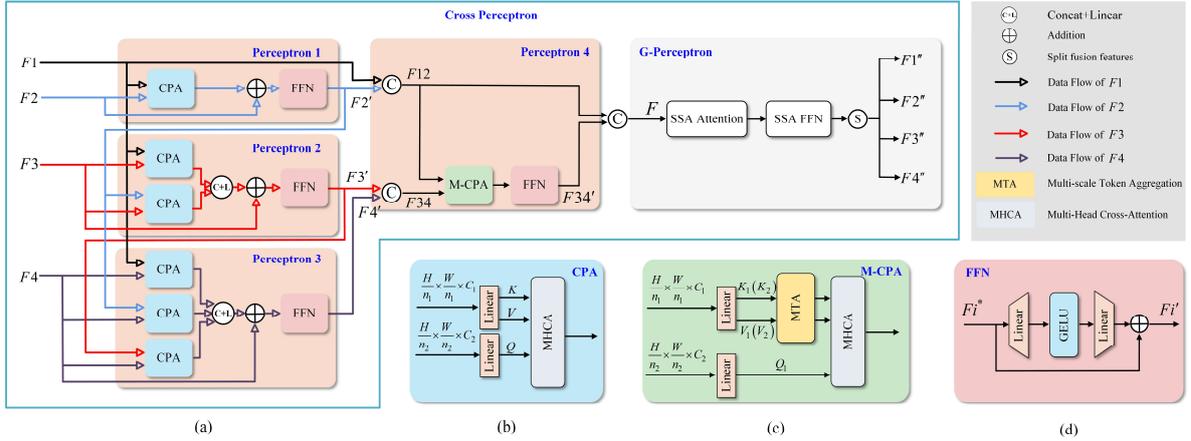

Fig. 2. The framework of the Cross Perceptron. (a) Overview architecture of the Cross Perceptron, including four Multi-scale Cross Perceptron modules (Perceptron 1 to Perceptron 4), and one Global Perceptron module. (b) The Cross Perceptron Attention (CPA). (c) The Modified-Cross Perceptron Attention (M-CPA). (d) The Feed Forward Network (FFN) of the Cross Perceptron.

TABLE I

THE SEQUENCE LENGTHS OF $Q$, $K$, AND $V$ IN EACH MULTI-SCALE CROSS PERCEPTRON MODULE (INCLUDING PERCEPTRON1 TO PERCEPTRON4) IN CROSS PERCEPTRON, AS WELL AS THE SEQUENCE LENGTHS IN G-PERCEPTRON

| Perceptor module | Lengths of $Q$ | Lengths of $K$ & $V$ |
|---|---|---|
| Perceptron 1 | $\frac{H \times W}{8 \times 8} \times 128$ | $\frac{H \times W}{8 \times 8} \times 128$ |
| Perceptron 2 | $\frac{H \times W}{16 \times 16} \times 256$ | $\frac{H \times W}{8 \times 8} \times 256$ $\frac{H \times W}{8 \times 8} \times 256$ |
| Perceptron 3 | $\frac{H \times W}{32 \times 32} \times 512$ | $\frac{H \times W}{8 \times 8} \times 512$ $\frac{H \times W}{16 \times 16} \times 512$ $\frac{H \times W}{16 \times 16} \times 512$ |
| Perceptron 4 | $\frac{H \times W \times (16+8)}{32 \times 32} \times 32$ | $\frac{H \times W \times (1+2)}{32 \times 32} \times 32$ $\frac{H \times W \times (1+2)}{32 \times 32} \times 32$ |
| G-Perceptron | $\frac{H \times W \times (64+32+16+8)}{32 \times 32} \times 32$ | $\frac{H \times W \times (1+2+4+8)}{32 \times 32} \times 32$ $\frac{H \times W \times (4+8+16+8)}{32 \times 32} \times 32$ |

$$Fi' = Fi^* + Linear(\sigma(Linear(Fi^*))), \qquad (5)$$

where, $Fi^*$ is the input to the FFN and $Fi'$ is the output of the FFN. Note that the output of the FFN has the same dimension as the input. Specifically, $F2'$, $F3'$, and $F4'$ correspond to $F2$, $F3$, and $F4$, respectively.

To further improve the performance, we also proposed a Modified-Cross Perceptron Attention (M-CPA) module to replace a CPA module in Perceptron 4. Similar to the SSA architecture, multi-scale $K$ and $V$ are generated by a multi-scale token aggregation (MTA) component within the M-CPA module, as shown in Fig. 2(c). The sequence length configurations of $Q$, $K$, and $V$ of the M-CPA module in Perceptron 4 are shown in Table I. In addition, each M-CPA input is obtained by concatenating the features of two adjacent scales. For example, $F12$ is obtained by concatenating $F1$ and $F2'$, while $F34$ is obtained by concatenating $F3'$ and $F4'$. The specific operation is defined as follows:

$$
\begin{aligned}
F12 &= Reshape\big(F1[-1,C_1]\big) \copyright Reshape\big(F2'[-1,C_1]\big); \\
F34 &= Reshape\big(F3'[-1,C_1]\big) \copyright Reshape\big(F4'[-1,C_1]\big),
\end{aligned}
\qquad (6)
$$

where $C_1$ is the depth of the channel.

Since each input of M-CPA is obtained by concatenating two features, the dimensions of $Q$, $K$, and $V$ are also adjusted accordingly. To minimize the computational complexity, the number of channels of $Q$, $K$, and $V$ is greatly reduced to 32, as shown in Table I.

As shown in Fig. 2(a), the channel depths of $F12$ and the output $F34'$ from Perceptron 4 are equivalent. These features are concatenated into a feature $F = (F12) \copyright (F34')$, which is then fed into the G-Perceptron module. The concatenated feature undergoes SSA Attention to compute the attention, followed by an SSA FFN to improve feature representability. The features representing global dependencies are then split based on their original order in terms of channel depth, resulting in four features: $F1''$, $F2''$, $F3''$, and $F4''$. These features correspond to the scales $F1$, $F2$, $F3$, and $F4$, respectively. Similarly, the sequence lengths of $Q$, $K$, and $V$ in the G-Perceptron are listed in Table I. The G-Perceptron module improves model performance by performing multi-headed attention over the entire global feature space, facilitating global dependency modeling, and integrating global features.

### B. Progressive Dual-branch Structure

To enhance the segmentation performance for medical images with finer tissue structures, such as retinal blood vessels in fundus images, we propose a Progressive Dual-branch Structure (PDBS) based on MCPA.

As shown in Fig. 3, our proposed PDBS consists of two branches, namely the Main Branch and the Fine Branch, both based on MCPA as the backbone network. The Main Branch is dedicated to the segmentation of the large-scale regions, while the Fine Branch focuses on the detection of fine details in medical images. To achieve accurate segmentation of fine-grained structures, we adopt the CNN module to replace the

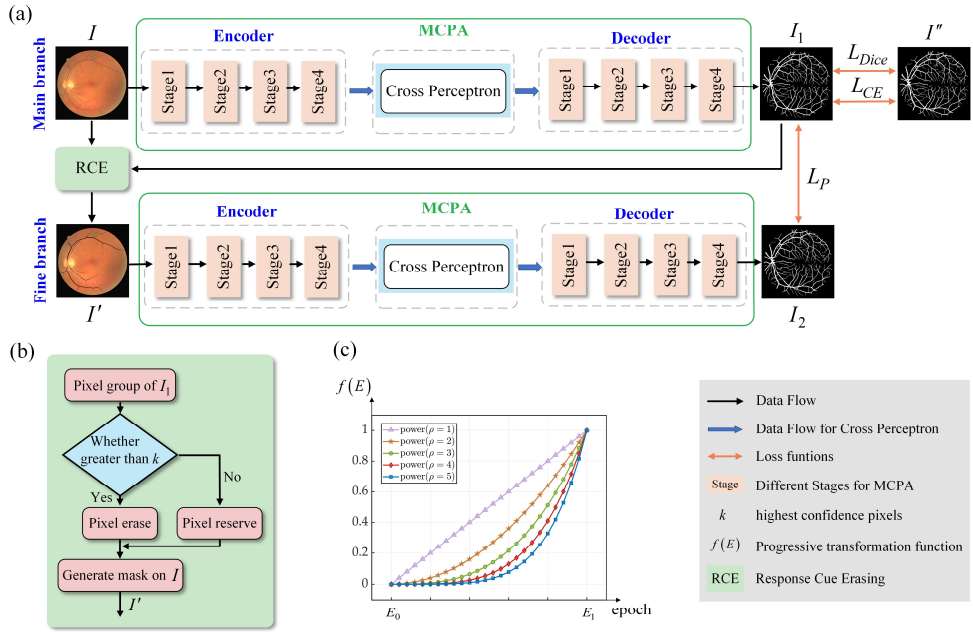

Fig. 3. The overall framework of the proposed Progressive Dual-branch Structure. (a) The network adopts two branches, where the Main Branch is used for extracting global features, while the Fine Branch further explores the potential for segmenting fine tissues and obtains the input for the Fine Branch through the Response Cue Erasing (RCE) module. (b) The operation of RCE module. (c) The curve of the monotonically increasing transformation function of the Progressive Regularization Loss, which varies with epoch, and the trend of the curves is different for different values of $\rho$.

Transformer modules in the encoder and decoder of MCPA. Specifically, we use CNN for both the encoder and the decoder, which is a common architectural choice seen in models such as UNet [7]. To seamlessly integrate of the Main Branch and the Fine Branch, we incorporate a Response Cue Erasing (RCE) module to establish a connection between the input of the Fine Branch and the output of the Main Branch. This allows the Fine Branch training to be easily guided by the results of the Main Branch.

The input image is denoted as $I$, while its corresponding ground truth label is denoted as $I''$, as shown in Fig. 3(a). After passing through the MCPA of the Main Branch, a segmentation map $I_1$ is generated. As shown in Fig. 3(b), the RCE module removes larger tissue structures (such as large blood vessels in retinal images) from the original image. The resulting image is then sent to the Fine Branch, which focuses its training focus on finer tissue structures (such as capillaries in retinal images). This process is detailed in:

$$I_1 = B_{Main}(I, \theta_1);$$
$$I_2 = B_{Fine}(R(I, I_1), \theta_2), \qquad (7)$$

where $R()$ represents the RCE module, $B_{Main}$ and $B_{Fine}$ correspond to the Main Branch and Fine Branch, respectively. $\theta_1$ and $\theta_2$ denote the parameters of the two networks. Specifically, we extract the top $k$ pixels with the highest confidence scores from the output result of the Main Branch. These pixels are considered as the large-scale tissue structure, which are easily identifiable throughout the region, while the remaining pixels are retained to form the input $I'$ for the Fine Branch. The modified image is then fed into the Fine Branch network to segment the finer tissue structures with more sophisticated features, resulting in a new feature map $I_2$.

Based on the loss described in [30], we incorporate regularization $Loss = \|I_1 - I_2\|_2$ to ensure consistency between the dual-branch, thereby increasing the robustness of the network. In the initial stage of training, when the model has not yet reached an optimized state, the segmentation results of the Main Branch may not be satisfactory. As a result, using its output to guide the training of the Fine Branch may not yield desirable results and may even have a negative impact, jeopardizing the achievement of an optimal network. To address this issue, we propose a progressive training strategy that gradually shifts the segmentation focus of the network from large-scale structures to fine structures. This is achieved by refining the aforementioned loss function and introducing a Progressive Regularization Loss (PR Loss, $L_P$) that progressively minimizes the discrepancy between the Main Branch and the Fine Branch:

$$L_P = f(E) \cdot (\|I_1 - I_2\|_2), \qquad (8)$$

where $f(E)$ denote a monotonically increasing transformation function that is depends on $E$ (current training epoch), which also satisfies $f(E_0)=0$ and $f(E_1)=1$. The specific equation for $f(E)$ can be formulated as follows:

$$f(E) = \begin{cases} 0, & E \leq E_0 \\ (\dfrac{E - E_0}{E_1 - E})^{\rho}, & E_0 < E < E_1 \\ 1, & E \geq E_1 \end{cases} \qquad (9)$$

TABLE II
COMPARISON OF THE SYNAPSE DATASET (AVERAGE DICE SCORE %, AVERAGE HAUSDORFF DISTANCE, AND DICE SCORE % FOR EACH ORGAN)
[KEY: **BEST**, <u>SECOND BEST</u>, <span style="color:red">**OUR PROPOSED METHOD**</span>]

| Methods | | Dice↑ | HD↓ | Aorta | Gallbladder | Kidney(L) | Kidney(R) | Liver | Pancreas | Spleen | Stomach |
|---|---|---|---|---|---|---|---|---|---|---|---|
| CNN | V -Net [31] | 68.81 | - | 75.34 | 51.87 | 77.10 | 80.75 | 87.84 | 40.05 | 80.56 | 56.98 |
| | R50 UNet [8] | 74.68 | 36.87 | <u>87.74</u> | 63.66 | 80.60 | 78.19 | 93.74 | 56.90 | 85.87 | 74.16 |
| | R50 Att-UNet [35] | 75.57 | 36.97 | 55.92 | 63.91 | 79.20 | 72.71 | 93.56 | 49.37 | 87.19 | 74.95 |
| Transformer | ViT [15] | 67.86 | 36.11 | 70.19 | 45.10 | 74.70 | 67.40 | 91.32 | 42.00 | 81.75 | 70.44 |
| | R50-ViT [15] | 71.29 | 32.87 | 73.73 | 55.13 | 75.80 | 72.20 | 91.51 | 45.99 | 81.99 | 73.95 |
| | TransUNet [15] | 77.48 | 31.69 | 87.23 | 63.13 | 81.87 | 77.02 | 94.08 | 55.86 | 85.08 | 75.62 |
| | SwinUnet [14] | 79.13 | 21.55 | 85.47 | 66.53 | 83.28 | 79.61 | 94.29 | 56.58 | 90.66 | 76.60 |
| | MISSFormer [16] | <u>81.96</u> | <u>18.20</u> | 86.99 | <u>68.65</u> | <u>85.21</u> | <u>82.00</u> | <u>94.41</u> | <u>65.67</u> | <u>91.92</u> | <u>80.81</u> |
| | <span style="color:red">**MCPA**</span> | **85.04** | **17.23** | **87.85** | **75.83** | **87.68** | **83.48** | **95.29** | **71.26** | **94.00** | **84.95** |

TABLE III
COMPARISON ON THE ACDC DATASET (AVERAGE DICE SCORE % AND AVERAGE HAUSDORFF DISTANCE IN MM, AS WELL AS DICE SCORE % FOR EACH ORGAN)
[KEY: **BEST**, <u>SECOND BEST</u>, <span style="color:red">**Our proposed method**</span>]

| Methods | Dice↑ | Ventricle (L) | Ventricle (R) | Myocardium |
|---|---|---|---|---|
| R50 UNet [8] | 87.55 | 94.92 | 87.10 | 80.63 |
| R50 Att-UNet [35] | 86.75 | 93.47 | 87.58 | 79.20 |
| ViT [15] | 81.45 | 92.18 | 81.46 | 70.71 |
| R50-ViT [15] | 87.57 | 94.75 | 86.07 | 81.88 |
| TransUNet [15] | 89.71 | 95.73 | 88.86 | 84.53 |
| SwinUnet [14] | 90.00 | **95.83** | 88.55 | 85.62 |
| MISSFormer [16] | <u>90.86</u> | <u>94.99</u> | <u>89.55</u> | <u>88.04</u> |
| <span style="color:red">**MCPA**</span> | **91.18** | <u>94.99</u> | **90.29** | **88.25** |

where the training process is divided into three stages based on a time threshold $[E_0, E_1]$. The hyperparameters $E_0$ and $E_1$ represent the start and end epochs of the progressive transition stage, respectively. Specifically, in the initial stage of training ($E \leqslant E_0$), i.e., $f(E_0)=0$, only the Main Branch is used to quickly obtain the optimized segmentation results for the large-scale tissue structure. Combined with the non-convexity of the loss function, the weights of the feature should be are slightly optimized during the transition stage of training ($E_0 < E < E_1$), rather than undergoing large changes. At the same time, the value of $f(E)$ transitions smoothly and continuously from 0 to 1, facilitating a gradual shift of the training focus to the more challenging task of fine tissue segmentation. Furthermore, a progressive hyperparameter $\rho$, based on a power law and illustrated in Fig. 3(c), $\rho = 1$ is introduced to further smooth the training trend. Notice that, $\rho = 1$ meaning $f(E)$ is the linear form. Finally, in the final stage ($E \geqslant E_1$), i.e., $f(E_1)=1$, the full Dual-branch Structure is used to further improve the segmentation results.

Therefore, the total loss $L_{total}$ is updated as follows, replacing (3) to:

$$L_{total} = \lambda L_{CE} + (1-\lambda)L_{Dice} + L_P . \quad (10)$$

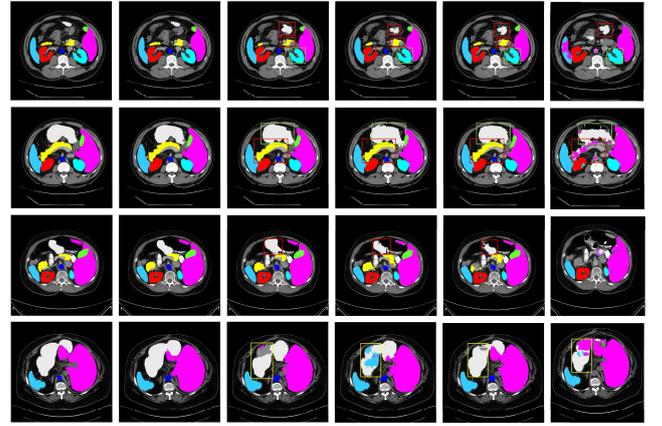

(a) Ground Truth (b) MCPA (c) TransUNet (d) SwinUnet (e) MissFormer (f) UNet

Fig. 4. Qualitative comparisons of different methods were made through visualization of the Synapse dataset. From left to right: (a) Ground Truth, (b) MCPA, (c) TransUNet, (d) SwinUnet, (e) MISSFormer, (f) UNet. The yellow boxes indicate organ confusion, green boxes indicate missing boundaries, and red boxes indicate organ deformation.

## III. EXPERIMENT

### A. Dataset and Implementation Details

This study uses four publicly available medical image datasets from different imaging modalities, including CT (Synapse), MRI (ACDC), fundus camera (DRIVE, CHASE_DB1, HRF, and OCTA (ROSE).

The Synapse multi-organ segmentation dataset (Synapse) comprises 30 cases with a total of 3779 axial abdominal clinical CT images. As described in [14], [15], the dataset provides annotations for 8 organs, including aorta, gallbladder, left kidney, right kidney, liver, pancreas, spleen, stomach. The dataset is divided into a training set of 18 samples and a test set of 12 samples. The Automated Cardiac Diagnosis Challenge dataset (ACDC) dataset consists of 100 samples obtained from different patients using an MRI scanner. For each MR image, segmentation annotations are available for the left ventricle

TABLE IV
Comparison on the DRIVE, CHASE DB1, HRF, and ROSE datasets (average Dice score% and AUC% of ROC)
[Key: **Best**, Second Best, **Our proposed method**]

| Methods | | DRIVE | | CHASE DB1 | | HRF | | ROSE | |
|---|---|---|---|---|---|---|---|---|---|
| | | Dice | AUC | Dice | AUC | Dice | AUC | Dice | AUC |
| CNN | UNet [7] | 81.42 | 97.55 | 80.74 | 98.42 | 78.56 | 97.87 | 71.16 | 92.18 |
| | UNet++ [10] | 81.92 | 98.12 | 81.34 | 98.35 | - | - | - | - |
| | Laddernet [36] | 82.02 | 97.93 | 80.31 | 98.39 | 77.31 | 97.65 | 78.11 | 94.39 |
| | CTF-net [37] | 82.41 | 97.88 | - | - | - | - | - | - |
| | Li *et al*. [30] | 82.61 | **98.43** | **81.67** | 98.35 | - | - | - | - |
| | OCTA-Net [29] | - | - | - | - | - | - | 76.97 | 94.53 |
| Transformer | TransUNet [15] | 81.18 | 97.48 | 81.19 | 97.83 | 79.50 | 98.13 | 77.09 | 93.21 |
| | SwinUnet [14] | 81.89 | 97.77 | 76.73 | 97.22 | 79.35 | 98.08 | 78.01 | 92.86 |
| | MISSFormer [16] | 81.90 | 97.91 | 81.05 | 98.52 | 79.27 | 98.05 | 77.70 | 93.14 |
| | **MCPA** | 82.12 | 98.17 | 81.25 | 98.49 | 79.64 | 98.20 | 78.20 | 94.50 |
| | **MCPA-CNN** | 82.87 | 98.06 | 81.19 | 98.57 | 79.61 | 98.18 | **78.58** | 94.62 |
| | **MCPA-CNN-PDBS** | **83.21** | 98.14 | 81.52 | **98.61** | **80.06** | **98.23** | 78.56 | **94.74** |

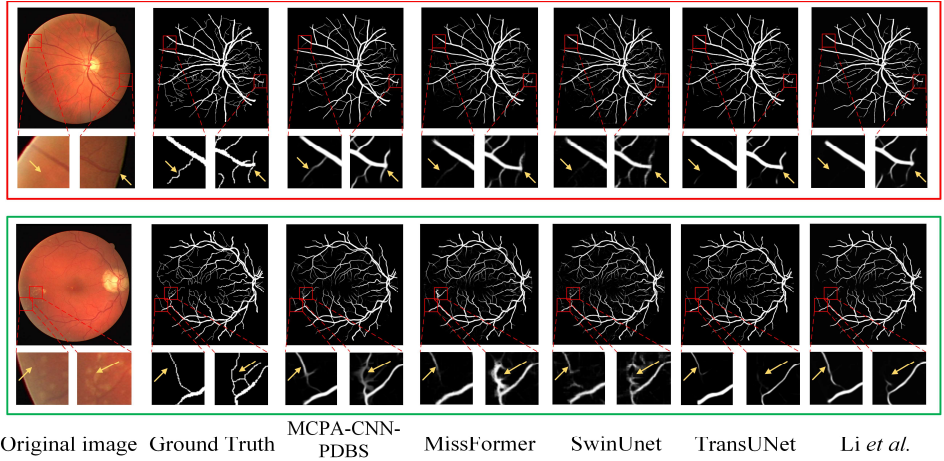

Fig. 5. Typical segmentation results of different methods on the DRIVE dataset, including Original image, Ground truth, MCPA-CNN-PDBS, MISSFormer, SwinUnet, TransUNet, Li *et al*.

Original image   Ground Truth   MCPA-CNN-PDBS   MissFormer   SwinUnet   TransUNet   Li *et al*.

(LV), right ventricle (RV), and myocardium (MYO). The dataset is further divided into 70 training samples, 10 validation samples, and 20 test samples, as reported in [14], [15]. The evaluation metrics differ between the datasets. For the Synapse datasets, the average Dice coefficient and the average 95% Hausdorff Distance (HD95) are used. For the ACDC datasets, the average Dice coefficient is used as the evaluation metric. In both datasets, the small SSA model pre-trained on ImageNet is used to initialize the MCPA network. The number of SSA Blocks in each stage of the encoder and decoder is set to [2, 4, 4, 1]. Following [16], the input image size is set to $224 \times 224$. The model is trained using the SGD optimizer for 400 epochs, with a batch size of 24, an initial learning rate of 0.04, a momentum of 0.9, and a weight decay of 1e-4.

The DRIVE dataset consists of 40 RGB retinal images with a resolution of $565 \times 584$ pixels. 20 images of them are assigned to the training set and the remaining 20 images are assigned to the test set [32]. The CHASE_DB1 dataset consists of 28 retinal images of left and right eyes, each with a resolution of $996 \times 960$ pixels. These images were collected from 14 school children and were divided into 20 training images and 8 test images [33]. The HRF dataset has a resolution of $3504 \times 2336$ and contains retinal images from 15 healthy subjects, 15 subjects with diabetic retinopathy, and 15 subjects with glaucoma. For this dataset, 36 images were used for training and 9 images were used for testing [34]. The Retinal OCT-Angiography vessel SEgmentation (ROSE) dataset [29] consists of 30 OCTA images of the subjects' superficial vessel complexes (SVC). These images have a resolution of $304 \times 304$ pixels. In this dataset, 22 images were used for training and 8 images were used for testing. We train the MCPA network using the Progressive Dual-branch Structure (PDBS). Training images were randomly sampling, each with a size of $48 \times 48$ pixels, moderate data augmentation techniques were applied. For model training, we used the Adam optimizer with

| SSA model size | Dice ↑ | HD ↓ | FLOPs (G) | Params (M) |
|---|---|---|---|---|
| Tiny [1, 2, 4, 1] | 82.26 | 25.13 | 8.46 | 27.13 |
| Small [1, 1, 1, 1] | 82.34 | 24.88 | 6.99 | 19.37 |
| Small [2, 4, 2, 1] | 84.25 | 18.77 | 9.33 | 26.16 |
| Small [2, 4, 4, 1] | **85.04** | **17.23** | 9.97 | 30.48 |
| Small [2, 4, 6, 1] | 83.32 | 20.62 | 10.61 | 34.79 |
| Small [2, 4, 8, 1] | 82.84 | 22.19 | 11.26 | 39.11 |
| Small [2, 4, 10, 1] | 82.63 | 22.82 | 11.90 | 43.43 |
| Small [2, 4, 12, 1] | 82.56 | 23.11 | 12.55 | 47.75 |
| Base [3, 4, 24, 2] | 82.43 | 25.14 | 17.07 | 77.48 |

| Achitecture | Backbone | G-Perceptron |
|---|---|---|
| Backbone | 84.05/21.71 | 84.44/20.98 |
| Perceptron 1 | 84.11/19.89 | 84.75/19.14 |
| Perceptron 1+2 | 84.16/19.62 | 84.87/19.08 |
| Perceptron 1+2+3 | 84.40/19.52 | 84.91/18.79 |
| Perceptron 1+2+3+4 | 84.45/18.59 | **85.04/17.23** |

default parameters, ran 80 epochs, set the batch size of 64 and an initial learning rate of 0.0008.

## B. Experiment results on Synapse and ACDC datasets

Our approach is compared to some of the best publicly available methods, including: Two 2D segmentation architectures based on CNN and Transformer were used for comparison on the Synapse and ACDC datasets. The Transformer-based methods used for comparison are TransUnet [15], SwinUnet [14], and MISSFormer [16], with their respective results reported in the corresponding papers. For a fair evaluation against the Transformer-based methods, the CNN-based methods are used, including V-Net [31], UNet with it's backbone replaced by ImageNet pre-trained ResNet-50, and Att-UNet [35].

Quantitative results: Table II and Table III show the results of the different methods on the Synapse and ACDC datasets. Compared to the CNN-based methods, the Transformer-based methods show superior segmentation performance, by emphasizing long-term dependencies and local context. For example, the CNN-based method in MISSFormer once achieved the highest performance. Compared to MISSFormer, the Multi-scale Cross Perceptron Attention Network (MCPA) shows an improvement of 3.08% and 0.97% in terms of Dice and HD95 evaluation metrics, respectively. In addition, MCPA achieves improved performance across eight different organs, with improvements ranging from 0.11% to 7.18% over previously established best results for Aorta to Stomach. In particular, significant improvements of 4% to 7% are observed for Gallbladder, Pancreas, and Stomach. On the ACDC dataset, MCPA improves the average Dice by 0.32% compared to the best performing MISSFormer.

Qualitative results: Fig. 4 shows the qualitative comparison results of different methods on the Synapse dataset. Note that CNN-based methods tend to produce similar local contexts, making it difficult to distinguish between organs and background. In addition, organs in close proximity often suffer from confusion (indicated by the yellow box), leading to a significant number of false negatives. In contrast, Transformer-based methods produce segmentations that are more comprehensive than CNN-based methods. However, they are also susceptible to problems such as missing boundaries (indicated by the green box), organ deformation (indicated by the red box), and confusion between different organ categories (indicated by the yellow box). These challenges can result from inadequate local modeling of transformer operations. In contrast, MCPA takes advantage of the effective fusion of local and global features to reduce overall organ deformation and confusion while better preserving organ boundary integrity.

## C. Experiment results on DRIVE, CHASE DB1, HRF, and ROSE datasets

For the retinal segmentation task, we compared 2D segmentation architectures based on CNN and Transformer, as detailed in Table IV. The CNN-based methods include UNet [7], UNet++ [10], LadderNet [36], CTF-Net [37], and the approach proposed by Li et al. using PA filters and RCE module [30]. These methods have demonstrated excellent performance in retinal segmentation tasks. In addition, we evaluated the performance of OCTA-Net separately on the ROSE dataset. The Transformer-based methods comprised TransUNet [15], SwinUnet [14], and MISSFormer [16]. The CNN-based MCPA was referred as MCPA-CNN. We used the Progressive Dual-branch Structure with MCPA-CNN as MCPA-CNN-PDBS. To provide a more comprehensive evaluation of vessel segmentation performance, we present the segmentation results and locally magnified results on the DRIVE dataset, as shown in Fig. 5.

As shown in Table IV, the MCPA method, based on a pure Transformer architecture, achieves Dice and AUC scores of 82.12 and 98.17 on the DRIVE dataset. Although it surpasses the Dice score of 0.22 and AUC score of 0.26 achieved by the best-performing Transformer-based MISSFormer method, there is still a performance gap when compared to some advanced CNN-based methods. In addition, MCPA-CNN improves the Dice score by 0.75, with only a marginal decrease in AUC of 0.11 compared to MCPA. Furthermore, the Dice score is also improved by 0.26 compared to the best CNN-based method proposed by Li et al. In addition, the MCPA-CNN-PDBS method based on MCPA-CNN improves the Dice score by another 0.34. These results indicate that replacing the Transformer module in our proposed MCPA model with a CNN module and adopting the Progressive Dual-branch Structure training method can lead to improved performance in segmenting fine structures in similar retinal images, as observed in Fig. 5.

In Table IV, we observe similar results for the CHASE_DB1, HRF dataset, and ROSE datasets. The proposed MCPA method shows superior performance compared to other transformer-based methods, and MCPA-CNN further improved this performance. Going one step further, MCPA-CNN-PDBS achieves even better results. For example, on the CHASE1 dataset, MCPA-CNN-PDBS achieves a Dice of 81.52 and an AUC of 98.61, outperforming Li *et al.*'s method by 0.26 in AUC. On the HRF dataset, MCPA-CNN-PDBS achieves a Dice of 80.06 and an AUC of 98.23, outperforming TransUNet by 0.56 in Dice and 0.1 in AUC. For the ROSE dataset, MCPA-CNN-PDBS achieves the highest AUC of 94.74 and the second highest Dice of 78.56, beating Laddernet by 0.45 in Dice and OCTA-Net by 0.21 in AUC.

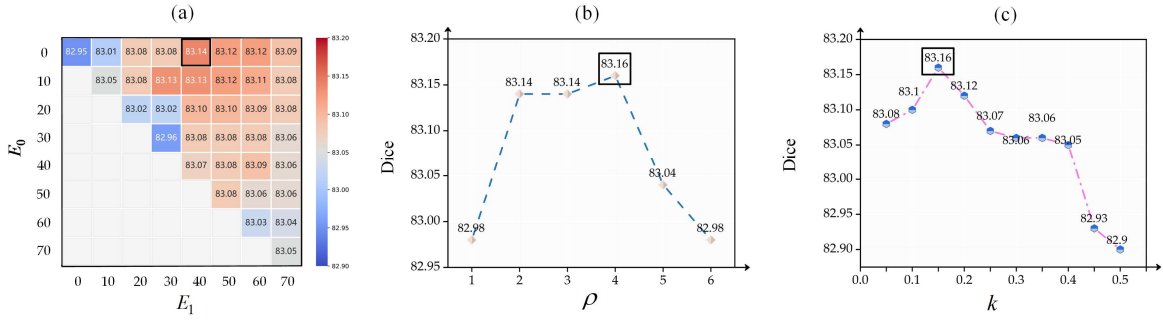

Fig. 6. Ablation studies on DRIVE Dataset. (a) Ablation studies of $[E_0,E_1]$ in PR Loss, (b) ablation studies of $\rho$ in PR Loss, (c) ablation studies of $k$ in RCE strategc.

## IV. ABLATION STUDIES

### A. Configurations of Model and Selection

According to reference [21], the encoder and decoder of the SSA model are divided into three different configurations: Tiny, Small, and Base, with pre-trained models on ImageNet. The number of SSA Blocks in each stage of the Tiny model is [1, 2, 4, 1], in the Small model is [2, 4, 12, 1], and in the Base model is [3, 4, 24, 2]. However, the acquisition of medical image data is often more difficult than the acquisition of natural data. herefore, for medical image segmentation tasks with small or medium-sized datasets, increasing the parameter capacity of similar model structures may not necessarily lead to improved segmentation results. This is because the model can be pushed from the underfitting region to the overfitting region [38], [39]. To determine the optimal capacity for a limited segmentation dataset when the network architecture remains similar, we performed ablation experiments on the Synapse dataset, as shown in Table V. To ensure a fair comparison, we used the pre-trained models of each model on ImageNet. Our results show that the original Tiny model achieved a Dice of 82.26, the Small model achieved a Dice of 82.56, and the larger Base model achieved a Dice of 82.43. We left the number of Blocks in the first two Stages of the Small model unchanged, i.e., 2 and 4, and mainly changed the number of Blocks in Stage 3, increasing it to 1, 2, 4, 6, 8, and 10. When the number of Stages is set to 4, the best Dice is 85.04 and the HD95 is 17.23. Therefore, we believe that choosing the Small model [2, 4, 4, 1] as the encoder will allow for more effective feature extraction.

### B. Impact of Cross Perceptron

We performed ablation experiments on the Synapse dataset to validate the effectiveness of the Cross Perceptron module in MCPA. We progressively incorporated four Multi-scale Cross Perceptron modules into the backbone and also conducted ablation studies on the Global module. As shown in Table VI, increasing the number of perceptrons resulted in a gradual improvement in the Dice and a decrease in the HD95. In addition, the inclusion of the G-Perceptron consistently improved performance. Optimal performance was observed when all four Perceptrons and the G-Perceptron were used.

### C. Progressive Dual-branch Structure

In the context of retinal segmentation, there are certain hyperparameters within the Progressive Dual-branch Structure that require optimization. These include the mask threshold $k$, the time threshold $[E_0,E_1]$, and the progressive hyperparameter $\rho$. We first performed ablation experiments on $E_0$ and $E_1$ on the DRIVE dataset, using a constant value of $\rho = 2$ and $k=0.15$. Additionally, we performed ablation studies on the parameters $E_0$ and $E_1$, ranging from the 10th epoch to the 70th epoch. The performance matrix is presented in Fig. 6(a). Notably, when $E_0$=0 and $E_1$=40 (represented by the black box), the Dice score achieved the highest performance of 83.14.

Furthermore, Fig. 6(b) shows the performance of different hyperparameters $\rho$ under the transformation function $f(E)$ during progressive training on the DRIVE dataset. For fairness, we set the time threshold $[E_0,E_1]$ to [0,40] and $k=0.15$. Our results show that the optimal Dice performance of 83.16 is obtained when $\rho$ is set to 4.

When performing the RCE operation, it is important to select an appropriate value for $k$. If $k$ is set too low, the erased region may not be clearly defined. Conversely, selecting a value that is too high may result in the loss of significant features. This in turn leads to a disproportionate loss of PR Loss between the Fine Branch and the Main Branch, biasing the segmentation towards fine vessels while neglecting the influence of $L_{CE}$ and $L_{Dice}$ on the features themselves. Thus, the choice of hyperparameters to determine the top $k$ highest confidence pixel positions has a significant impact on the experimental results. In Fig. 6(c), it is shown that on the DRIVE dataset, fixing the time threshold $[E_0,E_1]$ is fixed at [0,40] and $\rho = 4$, and $k=0.15$ provides a more effective Dice performance, reaching 83.16.

## V. CONCLUSION

This paper proposes a novel 2D medical image segmentation method called Multi-scale Cross Perceptron Attention Network (MCPA). The MCPA network an seamlessly integrate features across scales by using the Cross Perceptron to model both local details and global features. In addition, we propose the Progressive Dual-branch Structure to address the challenge of segmenting fine-structured tissues. It gradually shifts the focus of MCPA network training from whole-image segmentation to the more challenging task of segmenting fine pixel features. The proposed method has demonstrated improved segmentation performance on various medical images. In the future, we plan to explore the application of our feature fusion concept to other medical tasks, such as semantic segmentation, classification, and target detection of medical images in clinical settings.